\documentstyle[epsf]{elsart}

\def\bfnabla{\mbox{\boldmath $\nabla$}}
\def\bfSigma{\mbox{\boldmath $\Sigma$}}
\def\bfsigma{\mbox{\boldmath $\sigma$}}
\def\lQ{\Lambda_{\rm QCD}}
\newcommand{\nn}{\nonumber}
\newcommand{\be}{\begin{equation}}
\newcommand{\ee}{\end{equation}}
\newcommand{\bea}{\begin{eqnarray}}
\newcommand{\eea}{\end{eqnarray}}
\def\al{\alpha}
\def\als{\alpha_{\rm s}}
\def\siml{{\ \lower-1.2pt\vbox{\hbox{\rlap{$<$}\lower6pt\vbox{\hbox{$\sim$}}}}\ }} 

\begin{document}
\begin{frontmatter}
\begin{flushright}
\tt{CERN-TH/99-301\\ HEPHY-PUB 722/99\\ UB-ECM-PF 99/15\\ UWThPh-1999-60}
\end{flushright}
\vskip 1truecm
\title{The Heavy Quarkonium Spectrum at Order $m\als^5 \ln \als$}
\author {Nora Brambilla$^1$, Antonio Pineda$^2$, Joan Soto$^3$ and Antonio Vairo$^4$}
\address{$^1$ Institut f\"ur Theoretische Physik, 
     Boltzmanngasse 5, A-1090 Vienna, Austria}
\address{$^2$ Theory Division CERN, 1211 Geneva 23, Switzerland}
\address{$^3$ Dept. d'Estructura i Constituents de la Mat\`eria and IFAE, 
     U. Barcelona \\ Diagonal 647, E-08028 Barcelona, Catalonia, Spain}
\address{$^4$ Institut  f\"ur Hochenergiephysik, \"Osterr. Akademie d. Wissenschaften\\
     Nikolsdorfergasse 18, A-1050 Vienna, Austria}

\begin{abstract}
We compute the complete leading-log terms of the next-to-next-to-next-to-leading-order 
corrections to potential NRQCD. As a by-product we obtain the leading logs at $O(m\als^5)$ 
in the heavy quarkonium spectrum. These leading logs, when $\lQ \ll m\als^2$,  
give the complete $O(m\als^5 \ln \als)$ corrections to the heavy quarkonium spectrum. 
\end{abstract}

\vspace{1cm}
\end{frontmatter}

\newpage        

\pagenumbering{arabic}

\section{Introduction}
The last years have witnessed important progress in the theoretical
understanding of heavy-quark--antiquark systems near threshold through the use
of effective field theories \cite{Lepage,Manohar,pNRQCD,BS98,BV1,short,long} 
(see also \cite{KN,Labelle,pos,CMY} for related work in QED). The key point relies on the 
fact that, since the quark velocity $v$ is a small quantity, $v \ll 1$, a hierarchy of 
widely separated scales, $m$, $mv$, $mv^2$ ...,   
is produced in these systems, where $m$ is the heavy quark mass (hard scale). 
We will call $m v$ and $m v^2$ the soft and the ultrasoft (US) scale respectively.  
One can take advantage of this hierarchy by systematically integrating out the scales 
above the energies one wants to be described by the effective theory. 

After integrating out the hard scale, non-relativistic QCD (NRQCD) is obtained \cite{Lepage}. 
The Lagrangian of NRQCD can be organized in powers of $1/m$.  
The matching coefficients of NRQCD are (non-analytic) functions of $m$.
After integrating out the soft scale (proportional to the inverse of the size of the bound state)  
in NRQCD, potential NRQCD (pNRQCD) is obtained \cite{pNRQCD}. 
The Lagrangian of pNRQCD is organized in powers of $1/m$ and of the relative coordinate ${\bf r}$ 
(multipole expansion). The matching coefficients of pNRQCD are (non-analytic) functions of ${\bf r}$.

The integration of the degrees of freedom is done through a matching procedure 
(see \cite{Manohar,Match,pos,long} for details). The matching from QCD to NRQCD can always 
be done perturbatively since, by definition of heavy quark, $m\gg \lQ$ \cite{Manohar,Match}. 
The matching from NRQCD to pNRQCD can only be carried out perturbatively when
$mv \gg \lQ$. We will assume this to be so through this work. Therefore,
the matching coefficients both in NRQCD and in pNRQCD can be computed order by
order in $\als$. The non-analytic behaviour in $1/m$ appears through logs in 
the matching coefficients of the NRQCD Lagrangian:
$$
c \sim A\als \left( \ln{m \over \mu_h}+B \right),
$$
where $\mu_h$ denotes the matching scale between QCD and NRQCD. 
A typical matching coefficient of pNRQCD has the following
structure\footnote{We have assumed, since it will be so in the present work,
  that all the $\mu_h$ dependence cancels in the matching procedure. This is
  in general not true at higher orders due to the fact that there are terms in
  the quantum-mechanical perturbation series that also exhibit an explicit
  dependence on the cut-off $\mu_h$.}: 
$$
D \sim V({\bf r},{\bf p},{\bfsigma}_1,{\bfsigma}_2)\left(A^{\prime}\ln{m \over\mu_h}
+A^{\prime}\ln{\mu_h\,r}+B^{\prime}\ln{\mu\,r}+C\right),
$$
where $\mu$ denotes the matching scale between NRQCD and pNRQCD. $V$ denotes  
a function of {\bf r}, {\bf p} and  the spin of the particles, which is analytic in the two 
last operators but typically contains non-analyticities in  {\bf r}.
We see that the dependence on $\mu_h$ cancels here  
and that two, potentially large, logs appear in the pNRQCD matching coefficients: $\ln{mr}$ and $\ln{\mu r}$. 
They represent the leading-log corrections to the pNRQCD Lagrangian. 
In practice, the leading-log dependence on $\ln{mr}$ of the pNRQCD matching coefficients 
can be trivially extracted from the leading-log behaviour of the matching
between QCD and NRQCD at one loop, which is known, see \cite{Manohar,Match}. In order to obtain the leading-log dependence 
on $\ln{\mu r}$ of the pNRQCD matching coefficients, it is just sufficient to
know the behaviour  
of pNRQCD in the ultraviolet (UV) at next-to-leading order in the multipole
expansion. Moreover, everything can be worked out at some definite order in $1/m$. 
This will be discussed in greater detail below, where we will obtain the
leading-log dependence of the pNRQCD matching coefficients up to $O(1/m^2)$ and at leading order in the
multipole expansion. The leading-log running of the matching
coefficients at $O(1/m^0)$ has already been calculated in \cite{short}.

If we assume $\lQ \siml mv^2$ the leading-order solution to the heavy
quarkonium spectrum corresponds to a Coulomb-type bound state.
In a Coulomb-type bound state, $v \sim \als$ and all the terms of the type $\als/v$ have to be resummed 
exactly. The next-to-leading order (NLO) terms are corrections of $O(v,\als)$,
the next-to-next-to-leading order (NNLO) terms are corrections of $O(v^2,v\als,\als^2)$, and so on.
These corrections may get multiplied by parametrically large logs. 
The matching sketched above and exploited in the next sections is sufficient to obtain the leading-log corrections
of the NNNLO terms in the pNRQCD Lagrangian (there are not parametrically large logs at NLO and NNLO). 
Since results are now available for some observables up to NNLO order \cite{NNLO}, the leading-log corrections 
of the NNNLO terms in the pNRQCD Lagrangian are the first step towards the
complete evaluation of the heavy quarkonium spectrum at NNNLO order. 
It is one of the aims of the present work to set up the framework for an eventual full NNNLO calculation. 
For what concerns the relation between the heavy quarkonium mass and the pole mass our
calculation will allow us to obtain the complete $O(m\als^5\ln\als)$ corrections in the situation 
$mv^2 \gg \lQ$. In the more general situation, $\lQ \siml mv^2$, we will still be able to obtain all the 
$m\als^5\ln\displaystyle{m\als \over m}$ and $m\als^5\ln\displaystyle{m\als \over \mu}$ terms, 
where the $\mu$ dependence of the latter cancels against US contributions that now should be evaluated 
non-perturbatively. These US contributions have recently been computed perturbatively in the 
$\overline{{\rm MS}}$ scheme in Ref. \cite{KP} where the $\mu$ dependence can
already be read of. 
   
We distribute the paper as follows. In sections 2 and 3 we fix the effective field theory framework 
we are working in. In particular we write the pNRQCD Lagrangian in the equal mass case 
up to order $1/m^2$. In section 4 we perform the matching. In section 5 we apply the result to 
the quarkonium spectrum by computing the leading-log corrections at
$O(m\als^5)$. The last section is devoted to some comments and conclusions.

\section{NRQCD}
After integrating out the hard scale $m$, one obtains NRQCD \cite{Lepage}. 
Let us, first, write the most general Lagrangian (up to field redefinitions) up to $O(1/m^2)$:
\bea
\label{lNRQCD}
&&{\cal L}_{\rm NRQCD}= {\bar \Psi} \Biggl\{ i\gamma^0 D_0
+ \, {{\bf D}^2\over 2 m} + \, {{\bf D}^4\over 8 m^3}
+ c_F\, g {{\bf \bfSigma \cdot B} \over 2 m}
\\ \nonumber
&& \qquad
+ c_D \, g { \gamma^0 \left({\bf D \cdot E} - {\bf E \cdot D} \right) \over 8 m^2}
+ i c_S \, g { \gamma^0 {\bf \bfSigma \cdot \left(D \times E -E \times D\right) }\over 8 m^2} \Biggr\} \Psi
\\
&&
- {1\over 4} G_{\mu \nu} G^{\mu \nu} +{d_2\over m^2} G_{\mu \nu} D^2 G^{\mu \nu}
+{d_3 \over m^2} g f_{abc}G^a_{\mu\nu} G^b_{\mu\al} G^c_{\nu\al}
\nonumber \\
&&
\nn
+ {d_{ss} \over m^2} \psi^{\dag} \psi \chi^{\dag} \chi
+ {d_{sv} \over m^2} \psi^{\dag} {\bfsigma} \psi \chi^{\dag} {\bfsigma} \chi
+ {d_{vs} \over m^2} \psi^{\dag} {\rm T}^a \psi \chi^{\dag} {\rm T}^a \chi
+ {d_{vv} \over m^2} \psi^{\dag} {\rm T}^a {\bfsigma} \psi \chi^{\dag} {\rm T}^a {\bfsigma} \chi.
\eea
We have here $\Psi=\psi+\chi$, where $\psi$ is the Pauli spinor field that annihilates the fermion and $\chi$
is the Pauli spinor field that creates the antifermion, $\bfSigma = \displaystyle\left( \begin{array}{cc}
\bfsigma & 0 \\ 0 & \bfsigma \end{array} \right)$, $i D^0=i\partial_0
-gA^0$ and $i{\bf D}=i\bfnabla+g{\bf A}$. We have also included the ${\bfnabla}^4/m^3$
term in Eq. (\ref{lNRQCD}), since it will be, once the power counting is established, as important as
other $O(1/m^2)$ operators in the NRQCD Lagrangian. This Lagrangian is sufficient to obtain the pNRQCD Lagrangian up to, and including, the leading-log terms at NNNLO. 
One can see that this is so by following an argument analogous to Ref. \cite{pos}. 
The coefficients $c_F$, $c_D$, $c_S$, $d_2$ and $d_3$ can be found in Ref. \cite{Manohar} 
and $d_{ij}$ ($i,j=s,v$) in \cite{Match}. Since $d_2$ and $d_3$ do not contain $\ln m/\mu_h$ 
terms, they may be neglected for the present analysis.

\section{pNRQCD}
Integrating out the soft scale in (\ref{lNRQCD}) produces pNRQCD \cite{pNRQCD,long}. 
The pNRQCD Lagrangian reads as follows:
\begin{eqnarray}                         
& & {\cal L}_{\rm pNRQCD} =
{\rm Tr} \,\Biggl\{ {\rm S}^\dagger \left( i\partial_0 
- {{\bf p}^2\over m} +{{\bf p}^4\over 4m^3}
- V^{(0)}_s(r) - {V_s^{(1)} \over m}- {V_s^{(2)} \over m^2}+ \dots  \right) {\rm S}
\nonumber \\
&& \nonumber 
\qquad \qquad + {\rm O}^\dagger \left( iD_0 - {{\bf p}^2\over m}
- V^{(0)}_o(r) 
+\dots  \right) {\rm O} \Biggr\}
\nonumber\\
& &\qquad + g V_A ( r) {\rm Tr} \left\{  {\rm O}^\dagger {\bf r} \cdot {\bf E} \,{\rm S}
+ {\rm S}^\dagger {\bf r} \cdot {\bf E} \,{\rm O} \right\} 
+ g {V_B (r) \over 2} {\rm Tr} \left\{  {\rm O}^\dagger {\bf r} \cdot {\bf E} \, {\rm O} 
+ {\rm O}^\dagger {\rm O} {\bf r} \cdot {\bf E}  \right\}  
\nonumber\\
& &\qquad- {1\over 4} G_{\mu \nu}^{a} G^{\mu \nu \, a},
\label{pnrqcdph}
\end{eqnarray}
where we have explicitly written only the terms relevant to the analysis of the leading-log 
corrections at the NNNLO; S and O are the singlet and octet field respectively.  
All the gauge fields in Eq. (\ref{pnrqcdph}) are functions of the  
centre-of-mass coordinate and the time $t$ only.
For a more extensive discussion we refer the reader to Ref. \cite{long}.

The functions $V$ are the matching coefficients of pNRQCD. 
They typically have a non-analytic dependence on ${\bf r}$. Although in Eq. (\ref{pnrqcdph}) 
this dependence is not shown explicitly, the matching coefficients $V$ also depend 
on other parameters such as the mass (through logarithms), or operators such
as the
momentum and the spin (analytically). For the present purposes we can approximate $V_A$, $V_B$ 
and $V^{(0)}_o$ to their leading-order value in $\als$ (although the leading-log contributions are
known \cite{long}):
\be
V_A = V_B = 1;  \quad\qquad V^{(0)}_o(r) = \left({C_A\over 2} -C_F\right) { \als(r) \over r}.
\label{appr}
\ee
Let us now display the structure of the matching potentials $V^{(0)}_s$, $V_s^{(1)}$ and $V_s^{(2)}$, 
which are the relevant ones to our analysis.  

{\it 1) Order $1/m^0$}. From dimensional analysis, $V^{(0)}_s(r)$ can only
have the following structure: 
\begin{equation}
V^{(0)}_s(r) \equiv  - C_F {\alpha_{V_s}(r) \over r}.
\label{defpot0}
\end{equation}
{\it 2) Order $1/m$}. From dimensional analysis and time reversal $V^{(1)}_s$, can only have the 
following structure: 
\be
{V^{(1)}_s \over m} \equiv -{C_FC_A D^{(1)}_s \over 2mr^2}.
\label{V1}
\ee
{\it 3) Order $1/m^2$}. At the accuracy we aim, $V^{(2)}_s$ has the structure  
\bea
{V^{(2)}_s \over m^2} &=& 
- { C_F D^{(2)}_{1,s} \over 2 m^2} \left\{ {1 \over r},{\bf p}^2 \right\}
+ { C_F D^{(2)}_{2,s} \over 2 m^2}{1 \over r^3}{\bf L}^2
+ {\pi C_F D^{(2)}_{d,s} \over m^2}\delta^{(3)}({\bf r})
\nn
\\
& & + {4\pi C_F D^{(2)}_{S^2,s} \over 3m^2}{\bf S}^2 \delta^{(3)}({\bf r})
+ { 3 C_F D^{(2)}_{LS,s} \over 2 m^2}{1 \over r^3}{\bf L} \cdot {\bf S}
+ { C_F D^{(2)}_{S_{12},s} \over 4 m^2}{1 \over r^3}S_{12}(\hat{\bf r}),
\label{V2}
\eea
where $S_{12}(\hat{\bf r}) \equiv 3 \hat{\bf r}\cdot\bfsigma_1 \, \hat{\bf r}\cdot\bfsigma_2 
- \bfsigma_1\cdot\bfsigma_2$ and ${\bf S} = \bfsigma_1/2 + \bfsigma_2/2$. 
Note that ${\bf p}$ appears analytically in the matching potentials. 
The power on $n$ to which ${\bf p}$ appears in the potentials is constrained by the power in $1/m$. 
At order $1/m^2$, ${\bf p}$  can only appear at most at the square power\footnote{
This result follows from the knowledge of the operators present 
in the $1/m$ expansion in the underlying theory (NRQCD) and from the use of time reversal.}. 

The matching coefficients, $\alpha_{V_s}$, $D_s$, contain some $\ln r$
dependence once higher order corrections to their leading (non-vanishing) values are taken into account. 
In particular, we will write expressions like $\delta^{(3)}({\bf r})\ln r$. This is not 
a well-defined distribution and should be understood as $\displaystyle{1\over
  4\pi}{\rm reg} {1 \over r^3}$,  
which is the Fourier transform of $\ln 1/k$  (see \cite{pos} and references therein). 
Nevertheless, in order to use the same notation for all the matching
coefficients, and since it will be sufficient for the purposes of this paper,  
where we are only interested in the dependence on the logs of the matching 
coefficients, we will use the expression $\delta^{(3)}({\bf r})\ln r$, 
although it should always be understood in the sense given above.

Finally we note that the above representation of the potential can be related
to others found in the 
literature \cite{Gupta} by the use of the equations of motion and of the identity 
$$
- \left\{ {1 \over r},{\bf p}^2 \right\} + {1 \over r^3}{\bf L}^2 + 4\pi\delta^{(3)}({\bf r})
= - {1 \over r} \left( {\bf p}^2 + { 1 \over r^2} {\bf r} \cdot ({\bf r} \cdot {\bf p}){\bf p} \right).
$$
The ambiguity in the form of the potential is due to the freedom to
perform time-independent unitary field redefinitions, which do not change the
spectrum of the theory. One could try to fix this freedom by working with a minimal  
set of independent potentials (just in the same way as the matching coefficients of NRQCD are 
unambiguous for a given renormalization scheme once the Lagrangian is written 
in the minimal form, namely with no higher time derivatives). 
In principle, given a potential, one could obtain its minimal form by using field
redefinitions. This point will be elaborated elsewhere.

\section{pNRQCD matching coefficients}
In order to obtain the explicit expressions for the different matching coefficients in pNRQCD 
(i.e. ${\alpha}_{V_s}$, $D_s^{(1)}$, $D_s^{(2)}$, ...), one has to perform the
matching between NRQCD and pNRQCD. A detailed description of the procedure can
be found in \cite{pNRQCD,long,pos}. Let us first write our results for the matching 
coefficients up to $O(1/m^2)$. The matching coefficients read as follows:
\bea
&&{\alpha}_{V_s} =\alpha_{\rm s}(r)
\left\{1+\left(a_1+ 2 {\gamma_E \beta_0}\right) {\alpha_{\rm s}(r) \over 4\pi}\right.
\nonumber\\
&&\qquad\qquad +\left[\gamma_E\left(4 a_1\beta_0+ 2{\beta_1}\right)+\left( {\pi^2 \over 3}+4 \gamma_E^2\right) 
{\beta_0^2}+a_2\right] {\alpha_{\rm s}^2(r) \over 16\,\pi^2}
\left. + {C_A^3 \over 12}{\alpha_{\rm s}^3 \over \pi} \ln{\mu r}\right\},
\label{newpot0}\\ 
& &D^{(1)}_s=\alpha_{\rm s}^2(r)\left\{1+{2 \over 3}(4C_F+2C_A){\als \over \pi}\ln{\mu r} \right\},
\label{Ds1}\\ 
& &D^{(2)}_{1,s}=\alpha_{\rm s}(r)\left\{1+{4 \over 3}C_A{\als \over \pi}
\ln{\mu r} \right\},
\label{Ds2}\\ 
& &D^{(2)}_{2,s}=\alpha_{\rm s}(r),
\label{Ds22}\\ 
&& D^{(2)}_{d,s}= \alpha_{\rm s}(r)\left\{1+ {\als \over \pi}\left({2 C_F \over 3} 
+ {17 C_A\over 3}\right)\ln{m r} + {16\over 3}{\als \over \pi}\left({C_A\over 2} - C_F\right)\ln{\mu r}\right\},
\label{Dd2}\\ 
& &D^{(2)}_{S^2,s}=\als(r)\left\{1-{7 C_A \over 4}{\als \over \pi}\ln{m r} \right\} ,
\label{Dss2}\\ 
& &D^{(2)}_{LS,s}= \als(r)\left\{1-{2 C_A \over 3}{\als \over \pi}\ln{m r} \right\}, 
\label{DLs2}\\ 
& & D^{(2)}_{S_{12},s}=\als(r)\left\{1-C_A{\als \over \pi}\ln{m r} \right\}.
\label{Dsten2}
\eea

To obtain $\alpha_{V_s}$ with the above accuracy (Eq. (\ref{newpot0})), it is necessary to
perform the matching between NRQCD and pNRQCD (at $O(1/m^0)$) exactly at the two-loop level
and with leading-log accuracy at the three-loop level, i.e. to compute the
static potential to this order. The one-loop result was obtained in
Ref. \cite{1loop}, the two-loop one in Ref. \cite{twoloop} and the three-loop
leading log in \cite{short}. $\beta_n$ are the coefficients of the beta
function and the values of $a_1$ and $a_2$ can be found in Ref. \cite{twoloop}
(see also \cite{twoloop} for notation). 

\begin{figure}
\makebox[0.0cm]{\phantom b}
\put(162,9){$\underbrace{\hbox{~~~~~~~~~~~~~~~~~~~~~~~~~~~~~}}$}
\put(170,-12){$1/(E-V^{(0)}_o-{\bf p}^2/m)$}
\put(120,10){\epsfxsize=7truecm \epsfbox{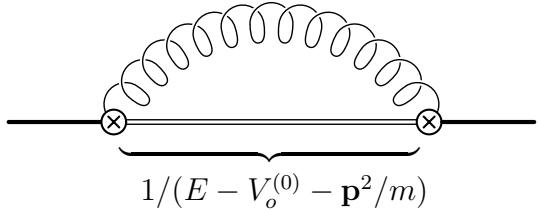}}
\caption{The UV divergences of this diagram in pNRQCD fix the $\mu$ dependence 
of $\alpha_{V_s}$, $D_s^{(1)}$, $D_s^{(2)}$, $Z_s^{(0)}$, $Z_s^{(1)}$ and $Z_s^{(2)}$.}
\vspace{3mm}
\label{US}
\end{figure}

For $D_s^{(1)}$ (Eq. (\ref{Ds1})) we need to perform the matching exactly 
at the one-loop level and to obtain the leading-log at the two-loop level. 
The one-loop matching has been done first (to our knowledge) in \cite{Duncan} 
(note that $D_s^{(1)}$ has no tree-level contributions). It gives 
$$
\delta D^{(1)}_s=\alpha_{\rm s}^2(r).
$$
The two-loop leading-log term is obtained from the UV behaviour of a next-to-leading 
order calculation in the multipole expansion to $O(1/m)$ in pNRQCD (see Fig. \ref{US}).

For the different $D_s^{(2)}$ terms we only need to compute the matching 
at tree level (along the same lines as in Ref. \cite{pos}) and the leading-log
at one loop. 
The leading dependence on $\ln mr$ is obtained by taking into account the
NRQCD matching coefficients in the vertices when matching to pNRQCD 
at tree level and by setting $\mu_h = 1/r$. The relevant contributions read as follows:
\begin{eqnarray*}
\delta D^{(2)}_{1,s}&=&\alpha_{\rm s}(r),\\ 
\delta D^{(2)}_{2,s}&=&\alpha_{\rm s}(r),\\ 
\delta D^{(2)}_{d,s}&=&\alpha_{\rm s}(r)(2+c_D-2c_F^2) +{1 \over \pi}\left[ d_{vs}+3d_{vv} 
+ {1 \over C_F}(d_{ss}+3d_{sv}) \right]\\
&\simeq& \alpha_{\rm s}(r)\left\{1+ {\als \over \pi}\left({2 C_F \over 3} 
+ {17 C_A\over 3}\right)\ln{m r}\right\},\\ 
\delta D^{(2)}_{S^2,s}&=&\alpha_{\rm s}(r)c_F^2 - {3 \over 2\pi C_F}(d_{sv}+C_F d_{vv})
\simeq \als(r)\left\{1-{7 C_A \over 4}{\als \over \pi}\ln{m r} \right\},\\
\delta D^{(2)}_{LS,s}&=&{\alpha_{\rm s}(r) \over 3}(c_S+2c_F)
\simeq \als(r)\left\{1-{2 C_A \over 3}{\als \over \pi}\ln{m r} \right\},\\
\delta D^{(2)}_{S_{12},s}&=&\alpha_{\rm s}(r) c_F^2
\simeq\als(r)\left\{1-C_A{\als \over \pi}\ln{m r} \right\}.
\end{eqnarray*}
The leading dependence on $\ln\mu r$ is obtained from the UV behaviour of a next-to-leading 
order calculation in the multipole expansion at $O(1/m^2)$ in pNRQCD (see Fig. \ref{US}).
It is worth noting that the spin-dependent matching potentials do not have 
$\ln \mu r$  contributions. This  follows from the fact that the vertices in Fig. \ref{US} 
have no spin structure. In fact, they are known with better accuracy than the one 
demanded here (see \cite{BV1,Gupta} and references therein).

In order to obtain the dependence on $\mu$ of the above matching coefficients,
it is sufficient,  
to compute the UV divergences of pNRQCD at the next-to-leading order in the multipole expansion 
and up to $O(1/m^2)$. Basically, we have to compute the diagram shown in Fig. \ref{US} (see \cite{long} 
for the Feynman rules in pNRQCD), say
$$
{i\over E  - {\bf p}^2/m - V_s^{(0)}} g^2 {C_F C_A \over 3} r^i 
\int {d^3k\over (2\pi)^3} k {i\over E - k - V^{(0)}_o - {\bf p}^2/m} r^i
{i\over E  - {\bf p}^2/m - V_s^{(0)}}
,
$$
where all the derivatives are assumed to act on the right. 
The divergences (the $\mu$ dependence) we find in this expression must cancel
against the $\mu$ dependence of the matching potentials in the expression:  
$$
-Z_s^{1/2} {i\over E - V_s^{(0)} - {\bf p}^2/m  - V_s^{(1)}/m - V_s^{(2)}/m^2} Z_s^{1/2\, \dagger}, 
$$
fixing the $\mu$ dependence of the different matching potentials 
listed in Eqs. (\ref{newpot0})--(\ref{Dsten2}). We also obtain the leading $\ln \mu r$ dependence 
of the singlet normalization factor $Z^{1/2}_s$ defined in \cite{long},  
\bea 
Z_{s,{\rm US}}^{1/2} &\equiv&\sqrt{N_c}\left(1+ Z_{s,{\rm US}}^{(0)} + {1\over  m\,r} Z_{s,{\rm US}}^{(1)} 
+ {1\over m^2} Z_{s,{\rm US}}^{(2)} {\bf p}^2 \right),
\label{Z}\\
Z_{s,{\rm US}}^{(0)}  &=& {1\over 4\pi} C_F C_A^2 \als^3  \ln \mu r,
\label{Z0}\\
Z_{s,{\rm US}}^{(1)} &=& {1\over \pi} \left( {4 \over 3} C_F^2 + 2 C_F C_A \right) 
\als^2 \ln \mu r,
\label{Z1}\\
Z_{s,{\rm US}}^{(2)} &=& {4\over 3\pi} C_F \als \ln \mu r,
\label{Z2}
\eea 
where again all the derivatives are assumed to act on the right.
As discussed in Sec. 3 the potential (in particular the matching coefficients given 
in Eqs. (\ref{newpot0})--(\ref{Dsten2})) is ambiguous due to the freedom to perform 
time-independent unitary field redefinitions. This ambiguity affects also the 
normalization factor, $Z^{1/2}_s$, so that terms in the potential can be
traded  
for terms in  $Z^{1/2}_s$ in a correlated way. Moreover, we note that, 
once the potential has been fixed, the normalization factor maintains at least a residual 
ambiguity, which amounts to $i {\cal O}$, $\cal O$ being a Hermitian operator 
that commutes with the singlet Hamiltonian, $\displaystyle{\bf p}^2/m + V_s^{(0)} + \dots$.

Finally, let us note that the order $1/m^0$ and $1/m$ terms seem to be
protected against hard contributions $O(\ln m)$. This has been explicitly
proved for the $1/m^0$ term (the static potential) at any finite order in $\als$ \cite{Appelquist}.

\section{Heavy quarkonium spectrum}
In this section we compute the heavy quarkonium spectrum up to, and including,
the leading logs at $O(m \als^5)$. Let us put ourselves in the situation $\lQ \siml mv^2$. 
The size of each term in the pNRQCD Lagrangian (\ref{pnrqcdph}) is
well-defined and can be evaluated as follows.  The relative momentum ${\bf p}$,
and the inverse relative coordinate $1/r$ have a size of $O(m\als)$. 
The time derivative has a size of $O(m\als^2)$. US gluon fields, derivatives acting
on it and the centre of mass momentum (in the rest frame, when entering in recoil corrections, 
due to the virtual emission of US gluons) have a size of the order of the next 
relevant scale ($m\als^2$ or $\lQ$). Therefore, in order to obtain the leading logs 
at $O(m \als^5)$ in the spectrum, $V^{(0)}_s$ has to be computed at $O(\als^4\ln)$,
$V_s^{(1)}$ at $O(\als^3\ln)$, $V_s^{(2)}$ at $O(\als^2\ln)$ and $V_s^{(3)}$ at $O(\als\ln)$. 
From the previous matching, we have obtained $V^{(0)}_s$, $V_s^{(1)}$ and
$V_s^{(2)}$ with the desired accuracy. Since terms of the type $O(\als\ln)$ can
not exist at tree level, $V_s^{(3)}$ does not need to be computed.  
The only term to be considered at $O(1/m^3)$ needs
to be $\als$-independent and it is the correction to the kinetic energy,  
$ -\displaystyle{\Delta^2 \over 4m^3}$. 

After these considerations we can obtain the $O(m\als^5)$ leading-log correction to the 
heavy quarkonium spectrum. From the potential--like terms (\ref{defpot0})- (\ref{V2}) 
we obtain the following correction to the energy:
\bea
&& \delta E_{n,l,j}^{\rm pot}(\mu) = E_n {\als^3 \over \pi} 
\left\{{C_A \over 3} \left[{C_A^2 \over 2} 
+4C_AC_F {1\over n(2l+1)}+2C_F^2\left({8 \over n(2l+1)} - {1 \over n^2}\right) \right]
\ln{\mu \over m\als} \right.
\nn \\ 
\nn && \qquad +{ C_F^2\delta_{l0} \over 3 n}\left(8\left[C_F-{C_A \over 2}\right]
\ln{\mu\over m\als} + \left[C_F+{17 C_A \over 2}\right]\ln{\als} \right)
\\ 
\label{energy1} 
&& \qquad \left. -{7 \over 3}{ C_F^2C_A \delta_{l0}\delta_{s1} \over n} \ln{\als}  
- {(1- \delta_{l0}) \delta_{s1} \over l(2l+1)(l+1)n}\,C_{j,l}{ C_F^2C_A \over 2} \ln{\als} \right\} \,,
\eea
where $E_n= - mC_F^2\als^2/(4n^2)$ and 
\begin{eqnarray*} 
C_{j,l} = \,\left\{
\begin{array}{ll}
\displaystyle{ -{(l+1)(4\,l-1) \over 2\,l-1}}&\ \ , \, j=l-1 \\
\displaystyle{-1}&\ \ ,\, j=l \\
\displaystyle{{l (4\,l+5)\over 2\,l+3}}&\ \ ,\, j=l+1.
\end{array} \right.
\end{eqnarray*}
The $\ln\als$ appearing in Eq. (\ref{energy1}) come from logs of the type 
$\displaystyle{\ln{1 \over m\, r}}$. Therefore, they can be deduced once the 
dependence on $\ln m$ is known. We have checked that our dependence on $\ln m$ coincides  
with the one obtained in Ref. \cite{Gupta}.  The $\mu$ dependence of Eq. (\ref{energy1}) 
cancels against contributions from US energies. It agrees with the $\mu$ dependence found 
in \cite{KP} for $l=0$. At the next-to-leading order in the multipole
expansion, the contribution 
from these scales reads
\begin{equation}
\delta^{\rm US} E_{n,l}(\mu) = -i{g^2 \over 3 N_c}T_F  \int_0^\infty \!\! dt 
\langle n,l |{\bf r} e^{it(E_n-H_o)} {\bf r}| n,l \rangle \langle {\bf E}^a(t) 
\phi(t,0)^{\rm adj}_{ab} {\bf E}^b(0) \rangle(\mu), 
\label{energyNP}
\end{equation}
where $H_o = \displaystyle{{\bf p}^2\over 2m} +V_o^{(0)}$ and $\mu$ is
the UV cut-off of pNRQCD. Then, the total correction to the energy is 
\be
\delta E_{n,l,j}= \delta^{\rm pot} E_{n,l,j}(\mu)+ \delta^{\rm US} E_{n,l}(\mu)\,.
\label{energytotal}
\ee

Different possibilities appear depending on the relative size of $\lQ$ with
respect to the US scale $m\als^2$. If we consider that $\lQ \sim m\als^2$ the gluonic correlator in 
Eq. (\ref{energyNP}) cannot be computed using perturbation theory. Therefore,
in a model independent approach, one can leave it as a free parameter and fix 
it with an experiment at some scale $\mu$ (since the running of
Eq. (\ref{energyNP}) with $\mu$ is known, one can then obtain its value at
another scale). Another possibility is to try to obtain it from lattice
simulations (see for instance \cite{lat}) or by using some models \cite{mar}. 

If we consider that $m\als^2 \gg \lQ$, Eq. (\ref{energyNP}) can be computed
perturbatively. Since $m\als^2$ is the next relevant scale, the effective role of
Eq. (\ref{energyNP}) will be to replace $\mu$ by $m\als^2$ (up to finite pieces that 
we are systematically neglecting) in Eq. (\ref{energy1}). Then Eq. (\ref{energytotal}) simplifies to 
\bea
&& \delta E_{n,l,j} = E_n {\als^3 \over \pi} \ln{\als} 
\left\{{C_A \over 3} \left[{C_A^2 \over 2} +4C_AC_F 
{1\over n(2l+1)}+2C_F^2\left({8 \over n(2l+1)} - {1 \over n^2}\right) \right] \right.
\nn
\\ 
&& \qquad \left. +{ 3 C_F^2\delta_{l0} \over n}\left[C_F+{C_A \over 2}\right] 
- {7 \over 3}{ C_F^2C_A \delta_{l0}\delta_{s1} \over n} 
- {(1- \delta_{l0}) \delta_{s1} \over l(2l+1)(l+1)n}\,C_{j,l}{ C_F^2C_A \over 2} \right\} \,.
\label{energy2}
\eea
Since in this situation one is assuming that $\displaystyle{\lQ \over m\als^2} \ll 1$, one
can expand on this parameter. Therefore, non-perturbative corrections can be
parameterized by local condensates. The leading and next-to-leading
non-perturbative corrections have been computed in the literature \cite{VL,P1}.

When working in the pole-mass scheme, it is expected that large corrections
will appear from the finite pieces at NNNLO. This problem seems to
be solved by using renormalon-based mass definitions or the like \cite{BBHLM}. 
In this case we could expect our calculation to provide an estimate of the magnitude 
of the NNNLO corrections.

\section{Conclusions}
We have computed the matching between NRQCD and pNRQCD and the heavy quarkonium spectrum 
at the leading-log NNNLO.  Within the pNRQCD effective field theory framework,
our results almost trivially follow from existing calculations. 
The results we achieved are important at least for $t$-$\bar t$ production and $\Upsilon$ physics.
In the first case the results we present are a step towards the goal of reaching 
a 100 MeV sensitivity on the top quark mass from the  $t$-$\bar t$ cross-section
near threshold to be measured at a future Next Linear Collider \cite{Beneke}. 
In the second case it will improve our knowledge on the $b$ mass. 

Our calculation paves the way for a complete NNNLO analysis of heavy quarkonium. 
Therefore, let us briefly comment here on what extra calculations are still required in order 
to obtain the pNRQCD Lagrangian that is relevant at NNNLO and the spectrum at $O(m\als^5)$. 
The contribution from the US scales have been computed in the $\overline{\rm MS}$ scheme 
at one loop \cite{KP}; also the matching between QCD and NRQCD is known at the one-loop level 
in the $\overline{\rm MS}$ scheme \cite{Manohar,Match}. Therefore, 
the missing calculations concern only the matching between NRQCD and pNRQCD. 
The static potential needs to be known at three loops, the $1/m$ corrections at two loops, 
the $1/m^2$ at one loop and the $1/m^3$ (if any) at tree level, all of them in the
$\overline{\rm MS}$ scheme so as to profit from the already known results.
\vspace{4mm}

{\bf Acknowledgements} 

N.B. acknowledges the TMR contract No. ERBFMBICT961714, A.P. the TMR contract No. ERBFMBICT983405, 
J.S. the AEN98-031 (Spain) and 1998SGR 00026 (Catalonia) and A.V. the FWF contract No. P12254;
N.B., J.S. and A.V. acknowledge the program ``Acciones Integradas 1999-2000'', project No. 13/99.
N.B. and A.V. thank the University of Barcelona for hospitality while part of this work 
was carried out.

\end{document}